# High responsivity, low dark current ultraviolet photodetector based on AlGaN/GaN interdigitated transducer

Peter F. Satterthwaite,[1,a)] Ananth Saran Yalamarthy,[2,a)], Noah A. Scandrette,[3] A. K. M. Newaz,[3] and Debbie G. Senesky[1,4,b)]

[1]*Department of Electrical Engineering, Stanford University, Stanford, California, 94305, USA*

[2]*Department of Mechanical Engineering, Stanford University, Stanford, California, 94305, USA*

[3]*Department of Physics and Astronomy, San Francisco State University, San Francisco, California, 94132, USA*

[4]*Department of Aeronautics and Astronautics, Stanford University, Stanford, California, 94305, USA*

An ultraviolet (UV) photodetector employing the two-dimensional electron gas (2DEG) formed at the AlGaN/GaN interface as an interdigitated transducer (IDT) is characterized under optical stimulus. The 2DEG-IDT photodetector exhibits a record high normalized photocurrent-to-dark current ratio (NPDR, 6 x $10^{14}$). In addition, we observe a high responsivity (7,800 A/W) and ultraviolet-visible rejection-ratio ($10^6$), among the highest reported values for any GaN photodetector architecture. We propose a gain mechanism to explain the high responsivity of this device architecture, which corresponds to an internal gain of 26,000. We argue that the valence band offset in the AlGaN/GaN heterostructure is essential in achieving this high responsivity, allowing for large gains without necessitating the presence of trap states, in contrast to common metal-semiconductor-metal (MSM) photodetector architectures. Our proposed gain mechanism is consistent with measurements of the scaling of gain with device channel width and incident power. In addition to high performance, this photodetector architecture has a simple two-step fabrication flow that is monolithically compatible with AlGaN/GaN high electron mobility transistor (HEMT) processing. This unique combination of low dark current, high responsivity and compatibility with HEMT processing is attractive for a variety of UV sensing applications.

___________________________

a) P. F. Satterthwaite and A. S. Yalamarthy contributed equally to this work

b) Author to whom correspondence should be addressed: dsenesky@stanford.edu

UV photodetectors have applications in diverse fields, including combustion flame detection, UV astronomy and satellite positioning.[1,2] GaN is an appealing materials platform for manufacturing UV photodetectors due to the maturity of GaN high electron mobility transistor (HEMT) fabrication technology, in addition to its ability to operate in high-temperature and radiation-rich environments.[3] In most applications, the ideal photodetector would exhibit a high responsivity to maximize the signal, in addition to a low dark current to minimize quiescent power. A performance metric which simultaneously captures these two values is the normalized photocurrent-to-dark current ratio (NPDR), defined as the ratio of responsivity to dark current, with units of 1/W.[4,5] Numerous photodetector architectures[2,6–14] have been demonstrated in GaN, with a broad range of reported responsivities and NPDRs, as summarized in Table 1. Table I additionally reports the UV-visible rejection-ratio, another important performance metric which determines the cross-sensitivity of the photodetector to visible light. A distinction can be drawn between devices with a responsivity that corresponds to less than 100% quantum efficiency[7,8] (0.29 A/W for 365 nm illumination), and those with a responsivity that exceeds this value.[6,9–13] In the latter category of photodetectors, which includes photoconductors,[10] phototransistors[9,15] and some metal-semiconductor-metal (MSM) photodetectors,[6,16] an internal gain mechanism must exist where each incident photon induces more than one electron in the conduction band. The gain ($G$) is defined as the ratio of charge carriers to the photon flux:[17]

$$G = \frac{hc}{e\lambda}\frac{I}{P} \tag{1}$$

where $I$ is the photocurrent, $\lambda$ is the wavelength of incident radiation and $P$ is the incident power. One particularly high gain photodetector architecture is the phototransistor; AlGaN/GaN phototransistors have been shown to achieve gains as high as 170,000.[9] Though these devices have a high gain, their fabrication requires the use of both n-and p-doped GaN.[9] Due to the high activation energies of all known acceptor-type dopants in GaN,[18] it is desirable to fabricate photodetectors which do not require doping. It is further desirable to fabricate devices compatible with HEMT fabrication, in order to leverage mature GaN technology and enable monolithic integration. Previous HEMT photodetectors,[11,19] which leverage the modulation of the AlGaN/GaN 2DEG sheet density under UV illumination, have been shown to have a high gain, however such devices also have high dark current, leading to some of the lowest NPDRs among reported devices (~$10^6$). Recent work[12,13] has shown that by introducing an intrinsic GaN channel between two 2DEG electrodes, a high NPDR can be achieved. Understanding the gain mechanism of such photodetectors is important for maturing this promising class of devices.

In this work, we present such a device with a record high NPDR ($6\times10^{14}$). In addition, the device has a responsivity (7,800 A/W), and UV-visible rejection-ratio ($10^6$) that are among the highest reported values. We propose that our device has a similar gain mechanism to that of a phototransistor, in a device architecture that is significantly simpler to fabricate, requiring two masks, and no doping. Evidence for this gain mechanism is provided by investigating the scaling of gain with channel length and incident power.

Devices were fabricated on an AlGaN/GaN-on-Si wafer (DOWA, Inc.) grown by metal-organic chemical vapor deposition (MOCVD). The III-nitride stack, illustrated schematically in Fig. 1a, consists of a 1.5-μm-thick strain management buffer structure and an active 1.2-μm-thick GaN layer grown on top of Si (111). Subsequent to the growth of the GaN layer, formation of the 2DEG was accomplished by growing an epitaxial stack consisting of a 1-nm-thick AlN spacer, 30-nm-thick $Al_{0.25}Ga_{0.75}N$ barrier layer and 1-nm-thick GaN capping layer. This wafer has a manufacturer specified 2DEG mobility of 1,400

cm$^2$/V-s and sheet density of 1x10$^{13}$ cm$^{-2}$ at room temperature. As shown in Fig. 1b, an array of 2DEG interdigitated transducers (2DEG-IDT) was fabricated by etching AlGaN 2DEG mesa electrodes. These 5-μm-wide 2DEG electrodes were separated by 5-μm-wide intrinsic (un-intentionally doped below 10$^{16}$ cm$^{-3}$) GaN buffer channels. Post mesa isolation, a standard Ti/Al/Pt/Au Ohmic metal stack was deposited, and activated with a 35 second, 850ºC anneal.[20] After this two-step process, the fabrication of the 2DEG IDT photodetectors was completed, however it should be noted that the wafer received subsequent processing for co-fabricated devices. In particular, standard MSM photodetectors were fabricated on the GaN buffer, using the same geometry as the 2DEG-IDT photodetectors. These devices (Pd-MSM) had Pd/Au (40 nm/10 nm) metal fingers in place of the AlGaN mesa electrodes for comparison studies.

Responsivity measurements were taken using a 365 nm UV lamp and semiconductor parameter analyzer (henceforth Setup I). All presented measurements were taken at room temperature. The results of these measurements for a characteristic 2DEG-IDT and Pd-MSM device under 1.5 mW/cm$^2$ optical power are shown in Fig. 2a. While both devices have a comparable, low dark current of ~10 pA, the 2DEG-IDT device has significantly higher photocurrent, corresponding to a responsivity of 2,500 A/W at 5 V, in contrast to the 0.78 A/W responsivity observed in the MSM photodetector at the same bias voltage. Though the 2DEG-IDT device has significantly higher responsivity, both devices have a responsivity which exceeds the 100% quantum efficiency limit (0.29 A/W for 365 nm illumination), indicating the presence of a gain mechanism in both devices.

To further probe the gain mechanism in the 2DEG-IDT device, we measured responsivity while varying the UV intensity across four orders of magnitude. These measurements are presented in Fig. 2b. Measurements with incident power above 0.010 mW/cm$^2$ were DC measurements performed with Setup I, and those with power below 0.010 mW/cm$^2$ were AC measurements, performed using a lock-in amplifier and a monochromated optical beam chopped at 200 Hz (henceforth Setup II). These data show that as power increases from 0.15 μW/cm$^2$ to 110 μW/cm$^2$, the responsivity increases dramatically by greater than 4 orders of magnitude, peaking at 7,800 A/W. Above 110 μW/cm$^2$, the responsivity decreases slightly to 2,500 A/W at 1.5 mW/cm$^2$ This increase in responsivity with increasing incident power, and subsequent saturation is consistent with previous reports of high gain GaN MSM photodetectors,[16] however opposite of the trend seen in phototransistors.[9]

Measurements of the transient response of the 2DEG-IDT device were conducted using Setup I with an optical chopper operating at 5 Hz. These measurements, presented in Fig. 2c, demonstrate rise and fall times of 32 ms and 76 ms, respectively, here defined as the time it takes the photocurrent to go from 10% to 90% of its final value. It is also observed that within a 200 ms window the photocurrent does not recover to its ~10 pA dark state value, indicating the presence of persistent photoconductivity,[21] common to AlGaN/GaN 2DEG photodetector devices. These rise and fall times are long relative to MSM photodetectors with no internal gain,[2] however they present a significant improvement on the 20 s rise and 60 s fall time observed in a photoconductor with comparable gain to our 2DEG-IDT device.[10]

Measurements of responsivity as a function of wavelength were also performed using Setup II. These data, shown in Fig. 2d, demonstrate a high UV-visible rejection ratio of 4 x 10$^6$, with a peak responsivity at ~362 nm. The broadband light source used in this measurement had a roughly constant intensity between 1 and 3 µW/cm$^2$ below the peak responsivity at 362 nm, and an increasing intensity between 3 and 28 µW/cm$^2$ as the wavelength increased from 362 to 430 nm. Because the responsivity of the 2DEG-IDT photodetector increases with incident power (as seen in Fig. 2), this measurement underestimates the true UV-visible rejection ratio. The band-pass nature of the spectral responsivity of this photodetector, where the responsivity decreases at wavelengths both below and above the GaN band gap (~365 nm), is consistent with previous reports of phototransistors,[9,15] indicating a similar gain mechanism in both devices.

We seek to explain the gain observed in both device architectures, in particular the extraordinarily high gain of 26,000 observed in the 2DEG-IDT. In order to understand the scaling laws of the gain mechanisms in MSM and 2DEG-IDT photodetectors, devices were fabricated where the intrinsic GaN channel length between the Pd and 2DEG electrodes was varied from 4 to 20 µm. Optical microscope images of the MSM and 2DEG-IDT structures are shown in the inserts of Fig. 3a and 3b respectively. Figures 3a and 3b show the gain vs. $1/L^2$, where $L$ is the spacing between the Pd and 2DEG electrodes respectively. These data show that while there is a clear linear relation, with zero y-intercept, between gain and $1/L^2$ for the 2DEG-IDT photodetectors (Fig. 3b), the same relation does not describe the gain vs. $1/L^2$ relation in the MSM photodetectors (Fig. 3a). This difference in length scaling, in addition to the vastly different response magnitudes, implies that different gain mechanisms are present in these two types of devices.

In the standard model of gain in photoconductors, which has been used by Kumar et al.[13] to describe the gain of an InAlN/GaN device with a similar architecture to this work, gain is due to carrier accumulation that is limited by recombination, leading to the following scaling law: $G = \mu_e \tau_r V / L^2$, where $\mu_e$ is the electron mobility, $\tau_r$ is the recombination time and $V$ is the applied voltage. Though this model reproduces the appropriate $1/L^2$ dependency, using a bulk mobility of 900 cm$^2$/V-s, which is appropriate for our unintentional doping concentration,[22] a recombination time of ~3 µs is required to fit our data. This value is three orders of magnitude larger than the previously reported ~6 ns minority carrier lifetimes in GaN.[23] Because the ~3 μs recombination time required to fit the standard model is unreasonable, a different gain mechanism is needed to describe the behavior of our 2DEG-IDT devices.

The proposed gain mechanism is schematically illustrated in the band diagram for a 2DEG-IDT photodetector presented in Fig. 3d. In the 2DEG-IDT device, an AlGaN/GaN valence band offset creates an energetic barrier which leads to hole accumulation at this interface. This hole accumulation leads to an electrostatic lowering of the energetic barrier for electrons in the 2DEG to escape the quantum well and enter the conduction band (dashed line in Fig. 3d). Using only the lowermost sub-band, the number of carriers per unit volume with sufficient energy to escape the 2DEG quantum well can be approximately written as:

$$n_e(\phi_b) \approx \frac{k_b T m^*}{\pi \hbar^2} \exp\left(-\frac{q(\phi_b - \Delta\phi_b)}{k_b T}\right) \tag{2}$$

where $m^*$ is the effective mass, $\phi_b$ is the energy separation between the Fermi level in the 2DEG and the top of the GaN conduction band in the dark state, and $\Delta\phi_b$ is the barrier lowering due to photon-induced hole accumulation (Fig. 3d). Assuming that all electrons with sufficient energy enter the conduction band, and that electron conduction between the 2DEG electrodes is due to drift, the gain is found to be:

$$G = \frac{n_e(\phi_b)}{n_{ph,tot}} \frac{\mu_e V}{L} \tag{3}$$

where $n_{ph,tot}$ is the total number of photons incident on the device. Evidence for a drift model comes from the fact that the photocurrent shown in Fig. 2a is approximately linear with applied voltage. This drift model can further explain the $1/L^2$ dependency of gain if $n_e(\phi_b) \propto F_{ph}$, where $F_{ph}$ is photon flux per unit area, because $n_{ph,tot} = F_{ph} L$. Using a mobility of 900 cm$^2$/V-s, the barrier height required to explain the observed photocurrents is ~150 meV. This order of magnitude is consistent with theoretical values for $\phi_b$, calculated to be ~100 meV using a commercial Schrödinger-Poisson solver.[24] The exponential nature of the model in Equation (2) further explains the observation in Fig. 2b, where increasing incident power increases responsivity, up to a point of saturation. At low incident powers, when $\phi_b - \Delta\phi_b \approx \phi_b$, small changes in $\Delta\phi_b$ lead to small changes in the number of conduction electrons. However, when $\phi_b - \Delta\phi_b$ is lowered at high incident powers, the same small change in $\Delta\phi_b$ will lead to an exponentially larger change in the number of conduction electrons. This continues until reaching a point of saturation where non-idealities, such as increased recombination in the channel and the AlGaN/GaN interface, lead to a divergence from this model. This gain mechanism is similar to that of a phototransistor, where minority carrier accumulation in the base lowers the base-emitter energetic barrier, allowing more majority carriers to be injected from the emitter.[17]

The band structure of an MSM photodetector, schematically illustrated in Fig. 3c has no barrier to hole conduction into the metal, in contrast to the 2DEG-IDT device. Thus, in order for spatial hole accumulation to occur, which is necessary to achieve a gain greater than unity,[25] a trap state must exist at the metal contact. The gain due to such a trap state would be limited by the density of trap states and the de-trapping time, thus the assumption $n_e(\phi_b) \propto F_{ph}$ is unlikely to hold. This explains the fact that gain in the Pd-MSM photodetector is not linearly proportional to $1/L^2$. Though trap states could exist at the AlGaN/GaN interface, the difference in the scaling of the gain with length in these devices indicates that trap states do not play the same role in determining the gain of the 2DEG-IDT device as they do in the Pd-MSM device.

In conclusion, we demonstrated a 2DEG-IDT photodetector with a record high NPDR ($6\times10^{14}$), in addition to a high responsivity (7,800 A/W) and UV-visible rejection-ratio ($10^6$). The observed 32/76 ms rise/fall times present significant improvements on the 20/60 s rise/fall times seen in a photoconductor with comparable gain. We argue that the gain mechanism in this device is similar to that of a phototransistor, where spatial hole accumulation leads to a lowering of the energy barrier for electrons entering the conduction band. This mechanism is consistent with the scaling of gain with incident power and device channel length. The simple, two-mask fabrication process further allows for monolithic integration of our device with AlGaN/GaN HEMTs, enabling on-chip integration of optical sensing systems using this material platform.

This work was supported by the National Science Foundation (NSF) Engineering Research Center for Power Optimization of Electro Thermal Systems (POETS) under Grant EEC-1449548. N.A.S and A.K.M.N. acknowledge the support from the National Science foundation grant ECCS-1708907. The devices were fabricated in the Stanford Nanofabrication Facility (SNF), which was partly supported by the NSF as part of the National Nanotechnology Coordinated Infrastructure (NNCI) under award ECCS-1542152.

## References

[1] H. Chen, K. Liu, L. Hu, A.A. Al-Ghamdi, and X. Fang, Mater. Today **18**, 493 (2015).

[2] E. Monroy, F. Omnès, and F. Calle, Semicond. Sci. Technol. **18**, R33 (2003).

[3] R.A. Miller, H. So, H.C. Chiamori, A.J. Suria, C.A. Chapin, and D.G. Senesky, Rev. Sci. Instrum. **87**, 095003 (2016).

[4] C.O. Chui, A.K. Okyay, and K.C. Saraswat, IEEE Photonics Technol. Lett. **15**, 1585 (2003).

[5] Y. An, A. Behnam, E. Pop, and A. Ural, Appl. Phys. Lett. **102**, 013110 (2013).

[6] S. Chang, M. Chang, and Y. Yang, IEEE Photonics J. **9**, 6801707 (2017).

[7] G.Y. Xu, A. Salvador, W. Kim, Z. Fan, C. Lu, H. Tang, H. Morkoç, G. Smith, M. Estes, B. Goldenberg, W. Yang, and S. Krishnankutty, Appl. Phys. Lett. **71**, 2154 (1997).

[8] T. Tut, M. Gokkavas, A. Inal, and E. Ozbay, Appl. Phys. Lett. **901**, 163506 (2007).

[9] W. Yang, T. Nohava, S. Krishnankutty, R. Torreano, S. McPherson, and H. Marsh, Appl. Phys. Lett. **73**, 978 (1998).

[10] L. Liu, C. Yang, A. Patanè, Z. Yu, F. Yan, K. Wang, H. Lu, J. Li, and L. Zhao, Nanoscale **9**, 8142 (2017).

[11] M.A. Kahn, M.S. Shur, Q. Chen, J.N. Kuznia, and C.J. Sun, Electron. Lett. **31**, 398 (1995).

[12] M. Martens, J. Schlegel, P. Vogt, F. Brunner, R. Lossy, J. Würfl, M. Weyers, and M. Kneissl, Appl. Phys. Lett **98**, 211114 (2011).

[13] S. Kumar, A.S. Pratiyush, S.B. Dolmanan, S. Tripathy, R. Muralidharan, and D.N. Nath, Appl. Phys. Lett. **111**, 251103 (2017).

[14] Z. Alaie, S.M. Nejad, and M.H. Yousefi, Mater. Sci. Semicond. Process. **29**, 16 (2014).

[15] M.L. Lee, J.K. Sheu, and Y.R. Shu, Appl. Phys. Lett. **92**, 053506 (2008).

[16] F. Xie, H. Lu, X. Xiu, D. Chen, P. Han, R. Zhang, and Y. Zheng, Solid. State. Electron. **57**, 39 (2011).

[17] S.M. Sze and K.K. Ng, *Physics of Semiconductor Devices* (John Wiley & Sons, Inc., Hoboken, NJ, USA, 2006).

[18] F. Mireles and S.E. Ulloa, Phys. Rev. B **58**, 3879 (1998).

[19] Z.H. Zaidi and P.A. Houston, IEEE Trans. Electron Devices **60**, 2776 (2013).

[20] M. Hou and D.G. Senesky, Appl. Phys. Lett. **105**, 081905 (2014).

[21] M.T. Hirsch, J.A. Wolk, W. Walukiewicz, and E.E. Haller, Appl. Phys. Lett. **71**, 1098 (1997).

[22] M. Shur, B. Gelmont, and M. Asif Khan, J. Electron. Mater. **25**, 777 (1996).

[23] Z.Z. Bandić, P.M. Bridger, E.C. Piquette, and T.C. McGill, Appl. Phys. Lett. **72**, 3166 (1998).

[24] S. Birner, T. Zibold, T. Andlauer, T. Kubis, M. Sabathil, A. Trellakis, and P. Vogl, IEEE Trans. Electron Devices **54**, 2137 (2007).

[25] Y. Dan, X. Zhao, and A. Mesli, arXiv:1511.03118 (2015).

TABLE I. Performance of various photodetector architectures demonstrated in GaN

| Detector type | Responsivity (A/W) | NPDR (1/W) | UV/VIS rejection-ratio | Bias Voltage (V) | References |
|---|---|---|---|---|---|
| MSM | 3.1 | $3\times10^{12}$ | $10^5$ | 10 | Chang et al.[6] |
| p-i-n diode | 0.15 | $2\times10^9$ | $10^3$ | 10 | Xu et al.[7] |
| Avalanche | 0.13 | $2\times10^{10}$ | $10^4$ | 20 | Tut et al.[8] |
| Phototransistor | 50,000 | $5\times10^{14}$ | $10^8$ | 3 | Yang et al.[9] |
| Photoconductor | 13,000 | $4\times10^8$ | $10^2$ | 1 | Liu et al.[10] |
| HEMT | 3,000 | $2\times10^6$ | $10^3$ | 10 | Khan et al.[11] |
| Meander | 10,000 | $8\times10^{12}$ | $10^4$ | 5 | Martens et al.[12] |
| Sliced-HEMT | 33 | $1\times10^{12}$ | $10^3$ | 5 | Kumar et al.[13] |
| 2DEG IDT | 7,800 | $6\times10^{14}$ | $10^6$ | 5 | This work |

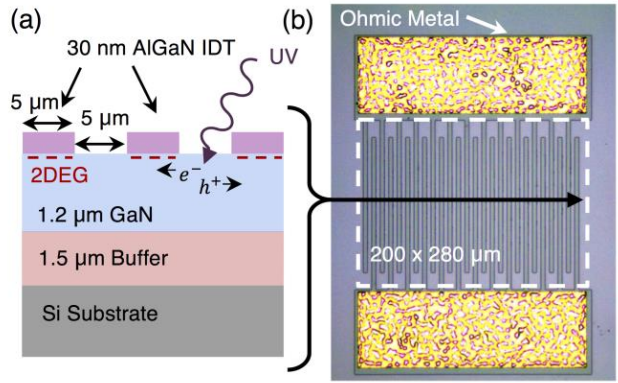

FIG. 1. (a) Schematic illustration of material stack.
(b) Optical microscope image of fabricated device.

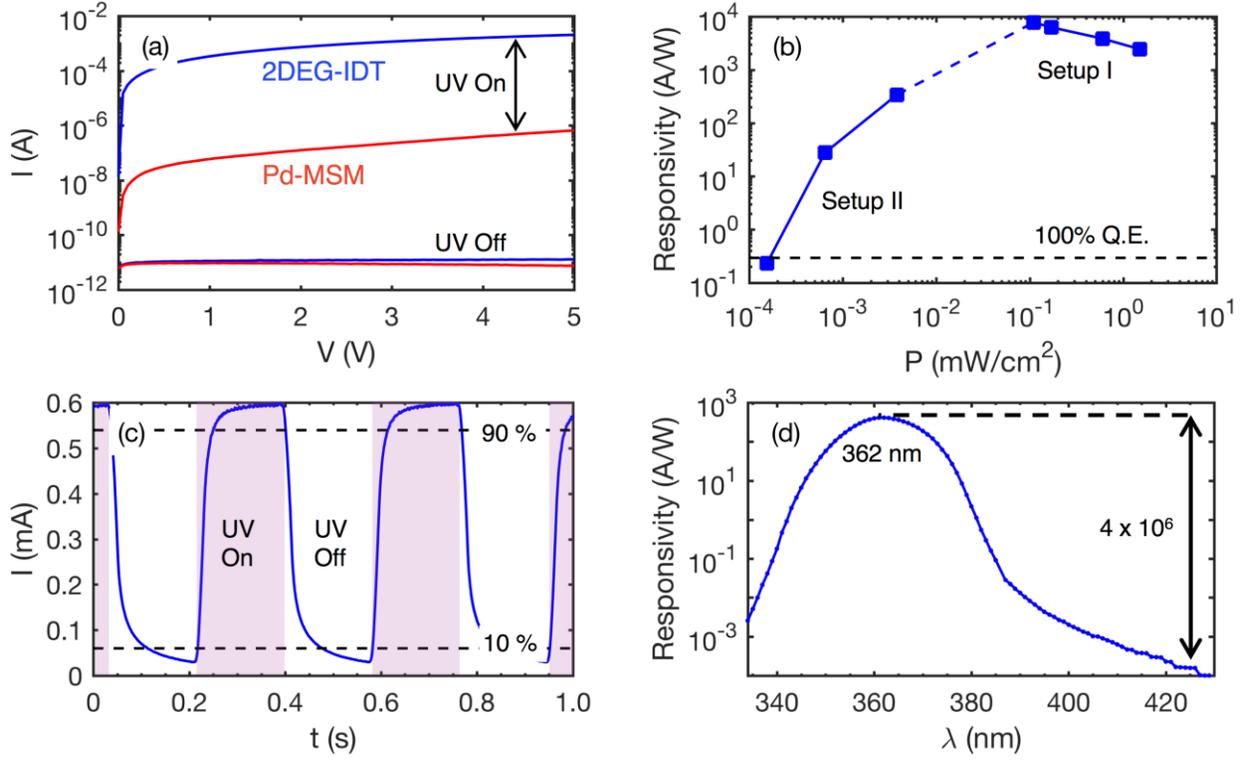

FIG. 2. (a) UV photoresponse of characteristic 2DEG-IDT and MSM devices exposed to 1.5 mW/cm$^2$ UV power. (b) Responsivity as a function of incident power, with the responsivity corresponding to 100% Q.E. labelled. Measurements below 0.010 mW/cm$^2$ were performed using Setup II, those above 0.010 mW/cm$^2$ were performed with Setup I. (c) Transient measurement of 2DEG-IDT response to 0.17 mW/cm$^2$ 365 nm illumination chopped at 5 Hz, measured in Setup I. (d) Responsivity vs. wavelength, measured in Setup II. Measurements in (b,c,d) were at a bias voltage of 5 V.

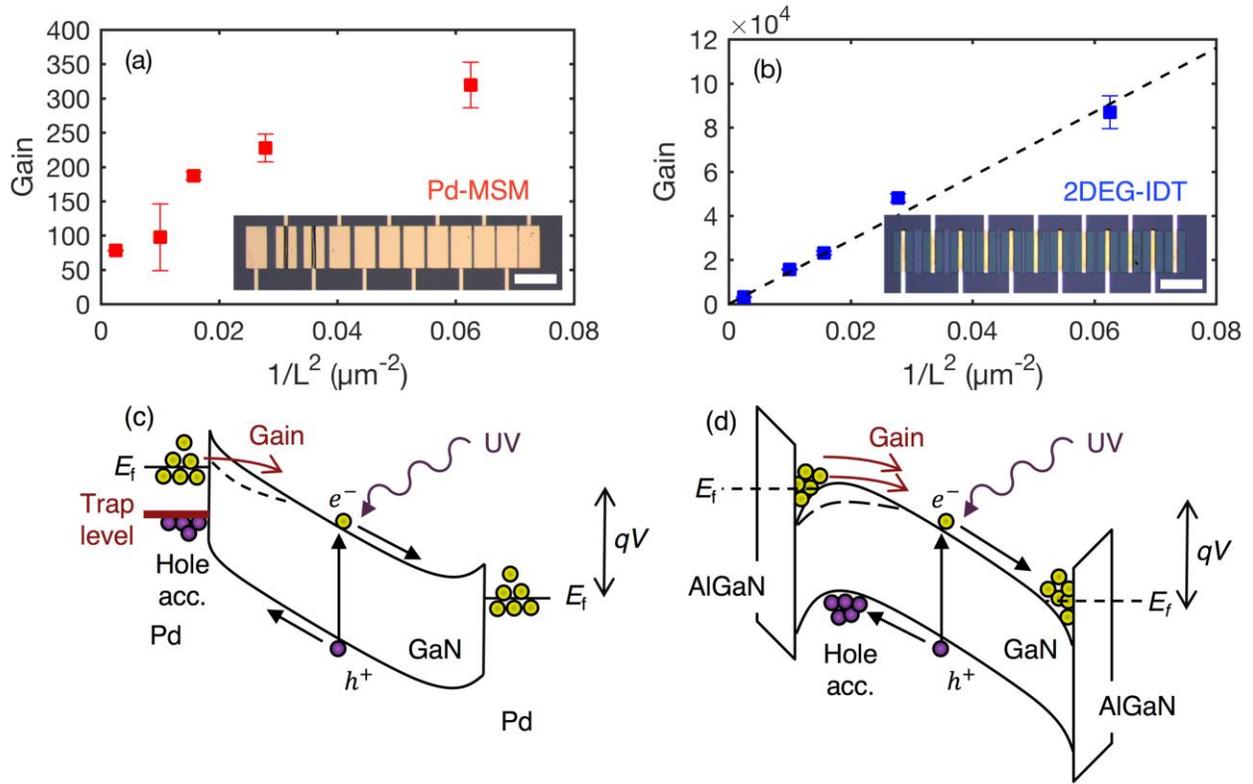

FIG. 3. Comparison of internal gain vs. length for (a) Pd-MSM and (b) 2DEG-IDT photodetectors. Error bars represent response biasing photodetector at $\pm 5$ V. Inserts show microscope images of fabricated devices with spacing of 4-20 μm, scale bars represent 100 μm. It is observed that there is a linear proportionality between gain and $1/L^2$ for the 2DEG-IDT device (black dashed line), but not for the Pd-MSM device. Band structures for the Pd-MSM and 2DEG-IDT photodetectors under applied bias are found in (c) and (d) respectively. Dashed lines in GaN conduction band represent electrostatic barrier lowering due to photon-induced hole accumulation.